\documentclass[prx,twocolumn,preprintnumbers,amsmath,amssymb,superscriptaddress]{revtex4}
\usepackage{epsfig}

\usepackage{graphicx}% Include figure files
\usepackage{dcolumn}% Align table columns on decimal point
\usepackage{bm}% bold math
\usepackage{float}
\usepackage{hyphenat}
\usepackage[dvipsnames]{xcolor}
\usepackage{physics}
\usepackage[T1]{fontenc}
\usepackage{gensymb}
\usepackage[utf8]{inputenc}
\usepackage{amsmath}
\usepackage{amsthm}
\usepackage[normalem]{ulem}

\DeclareUnicodeCharacter{2212}{\ensuremath{-}}

\newcommand{\heriotwatt}{Institute of Photonics and Quantum Sciences, SUPA, Heriot-Watt University, Edinburgh EH14 4AS, UK}

\begin{document}

\title{Twist-Engineered Nonlinearity in Two-Dimensional Crystals for Tailored Quantum Light}

\author{Dylan Mcleod}
\affiliation{\heriotwatt}

\author{Fabrizio Chiriano}
\affiliation{\heriotwatt}

\author{Francesco Graffitti}
\affiliation{\heriotwatt}

\author{Alessandro Fedrizzi}
\affiliation{\heriotwatt}

\author{Brian D. Gerardot}
\affiliation{\heriotwatt}

\author{Mauro Brotons-Gisbert}
\email{m.brotons_i_gisbert@hw.ac.uk}
\affiliation{\heriotwatt} 

\date{\today}

\begin{abstract}

Van der Waals (vdW) materials enable nonlinear-optical engineering with unprecedented resolution: their strong second-order susceptibilities ($\chi^{(2)}$) and twist-tunable interlayer symmetry allow the effective nonlinearity to be shaped continuously, rather than through binary $\pm\chi^{(2)}$ domain inversion as in bulk ferroelectrics. Here, we show that twist-angle domain engineering exploits this continuous degree of freedom to reconstruct target longitudinal nonlinearity profiles with high fidelity. Using spontaneous parametric down-conversion (SPDC) as a benchmark, we demonstrate that twist-engineered vdW crystals yield significantly improved approximations of target phase-matching functions and correspondingly higher single-photon purities, particularly in compact devices where fabrication constraints limit conventional approaches. We further show that this framework remains effective in experimentally relevant vdW materials and demanding non-degenerate wavelength regimes involving mid-infrared photons. More broadly, the ability to continuously and locally program $\chi^{(2)}$ establishes a general framework for tailoring a wide range of SPDC properties—including absolute brightness, joint spectral amplitude structure, signal–idler frequency separation, and temporal wavepacket shape—beyond what is accessible in conventional nonlinear crystals. These results position vdW heterostructures as a powerful platform for engineered quantum light sources and open new opportunities for nonlinear-optical devices shaped with monolayer thickness scale.

\end{abstract}

\maketitle
\section{Introduction}
The ability to engineer the longitudinal nonlinearity profile of an optical medium is essential for tailoring a wide range of properties of photons generated through nonlinear wave-mixing processes. Precise control of the crystal nonlinearity and therefore the phase-matching function (PMF) enables optimization of performance metrics that are critical across both classical and quantum photonics, including high-efficiency frequency conversion \cite{kumar1990quantum, Hugo2021quantum}, spectrally pure single-photon generation \cite{branczyk2010engineered, dixon2013spectral, dosseva2016shaping, pickston2021optimised}, and time frequency mode generation \cite{graffitti2020direct}. In bulk nonlinear crystals \cite{kwait1995new,pelton2004bright,lim2008stable, Rangarajan2009optimizing}, techniques such as periodic poling \cite{yamada1993first}, apodization \cite{heese2012role}, and various custom poling regimes \cite{branczyk2010engineered, dosseva2016shaping, okano2016resolution, Weiss2025nonlinear} have long been established to reshape the effective nonlinear response, yielding crystals with tailored nonlinearities approximating desired target functions. However, these approaches are inherently constrained by the binary choice of domain orientation and by fabrication limitations of macroscopic ferroelectric crystals. While shorter domains in the range of approximately 200-500 nm have been demonstrated using advanced fabrication techniques \cite{krasnokutska2021submicron,kuo2023photon}, such structures are typically confined to shallow surface layers or waveguide geometries and have not yet scaled to bulk interaction lengths, leaving bulk nonlinear interactions practically limited to micron-scale domain engineering.

Van der Waals (vdW) crystals offer a new paradigm for nonlinear optical engineering \cite{du2024nonlinear}. Their atomically thin nature and exceptionally large second-order nonlinear susceptibilities \cite{zograf2024combining}, often enhanced by excitonic resonances \cite{wang2015giant}, provide a powerful platform for tailoring nonlinear interactions at the nanoscale. In addition, their compatibility with arbitrary stacking and seamless integration into photonic chips \cite{xu2022towards} and fibers \cite{lin2026nonlinear} enables device architectures that are difficult to realize in conventional bulk materials. Spontaneous parametric down-conversion (SPDC) has already been experimentally demonstrated in a range of non-centrosymmetric, atomically layered vdW materials, including NbOCl$_2$ \cite{guo2023ultrathin,kallioniemi2025van}, rhombohedral polytypes of MoS$_2$ (3R-MoS$_2$) \cite{weissflog2024tunable} and WS$_2$ (3R-WS$_2$) \cite{feng2024polarization}, as well as boron nitride (rBN) \cite{liang2025tunable,lin2026nonlinear}.
Beyond reproducing quasi-phase-matching strategies previously developed in bulk systems \cite{qi2024stacking,trovatello2025quasi,tang2024quasi,qin2024interfacial}, vdW materials introduce an additional and fundamentally new degree of freedom: the interlayer twist angle. Controlled twisting can break and reconfigure crystal symmetries \cite{yao2021enhanced,tang2024quasi} and enable “twist-phase matching” through the nonlinear geometric phase associated with interlayer rotation \cite{hong2023twist,lin2026nonlinear}. The interlayer twist angle can therefore be exploited to modulate the effective second-order nonlinear susceptibility $\chi^{(2)}$ with atomic-scale precision, providing a level of spatial resolution and tunability inaccessible in conventional bulk ferroelectrics. As a result, vdW crystals open the possibility of realizing near-arbitrary, high-fidelity shaping of the effective nonlinearity, enabling more direct control over the properties of photons generated through nonlinear optical processes.

Beyond enabling high-resolution control of the effective nonlinearity, vdW crystals also offer access to wavelength regimes that are challenging for conventional bulk nonlinear platforms. In particular, strongly non-degenerate SPDC—where the signal and idler photons occupy widely separated spectral regions, such as the near- and mid-infrared—is often constrained in bulk crystals by material transparency, phonon absorption, dispersion asymmetry, and limited phase-matching flexibility \cite{boyd2008nonlinear, kuo2006optical}. The combination of large $\chi^{(2)}$ responses, atomic-scale thickness, and twist-angle–programmable nonlinearity in vdW heterostructures provides a promising route to overcoming these limitations, motivating the exploration of twist-engineered SPDC beyond near-degenerate and telecom-centric regimes.

Photon purity in SPDC provides a natural benchmark for assessing the impact of longitudinal nonlinearity engineering, particularly in regimes where spectral correlations are exacerbated by dispersion or wavelength asymmetry, as it is both experimentally accessible and directly determined by the crystal PMF. In periodically poled bulk crystals, SPDC photon pairs are typically generated with strong spectral correlations, which reduce the heralded single-photon purity. Custom-poled and apodized bulk crystals can mitigate these correlations by approximating Gaussian nonlinearity profiles \cite{branczyk2010engineered,tambasco2016domain,dosseva2016shaping,graffitti2017pure}; however, the fidelity of such approximations is fundamentally constrained by the binary nature of ferroelectric domain orientation. 

In vdW systems, by contrast, twist-angle domain engineering provides a qualitatively different approach to tailoring the effective nonlinearity profile, with spatial resolution ultimately set by the atomic lattice. This continuous degree of freedom suggests that vdW systems can achieve photon-purity levels beyond those accessible with bulk ferroelectric techniques, even in regimes where strong dispersion asymmetry or non-degenerate wavelengths would otherwise limit performance, particularly in compact devices where binary poling imposes a fundamental constraint.

In this work, we: (i) introduce a twist-angle domain-engineering algorithm that exploits the continuous tunability of $\chi^{(2)}$ in vdW crystals; (ii) use it to reconstruct Gaussian longitudinal nonlinearity profiles with high fidelity; (iii) evaluate the resulting single-photon purities in spontaneous parametric down-conversion (SPDC), including in strongly non-degenerate wavelength configurations; and (iv) benchmark the achievable performance as a function of domain width, crystal length, and twist-angle step size.
Together, our results highlight the unique advantages of vdW crystals for nonlinear optical engineering, establishing them as a promising platform for high-purity photon-pair generation. More broadly, the same framework naturally extends to the independent control of a wider range of SPDC properties — including absolute brightness, detailed joint spectral amplitude shaping (e.g. asymmetric or multi-lobed spectra), signal–idler frequency separation, and temporal wavepacket engineering — offering a versatile route toward photon-pair sources tailored to specific experimental requirements.

\begin{figure*}
    \begin{center}
    	\includegraphics[scale= 0.7]{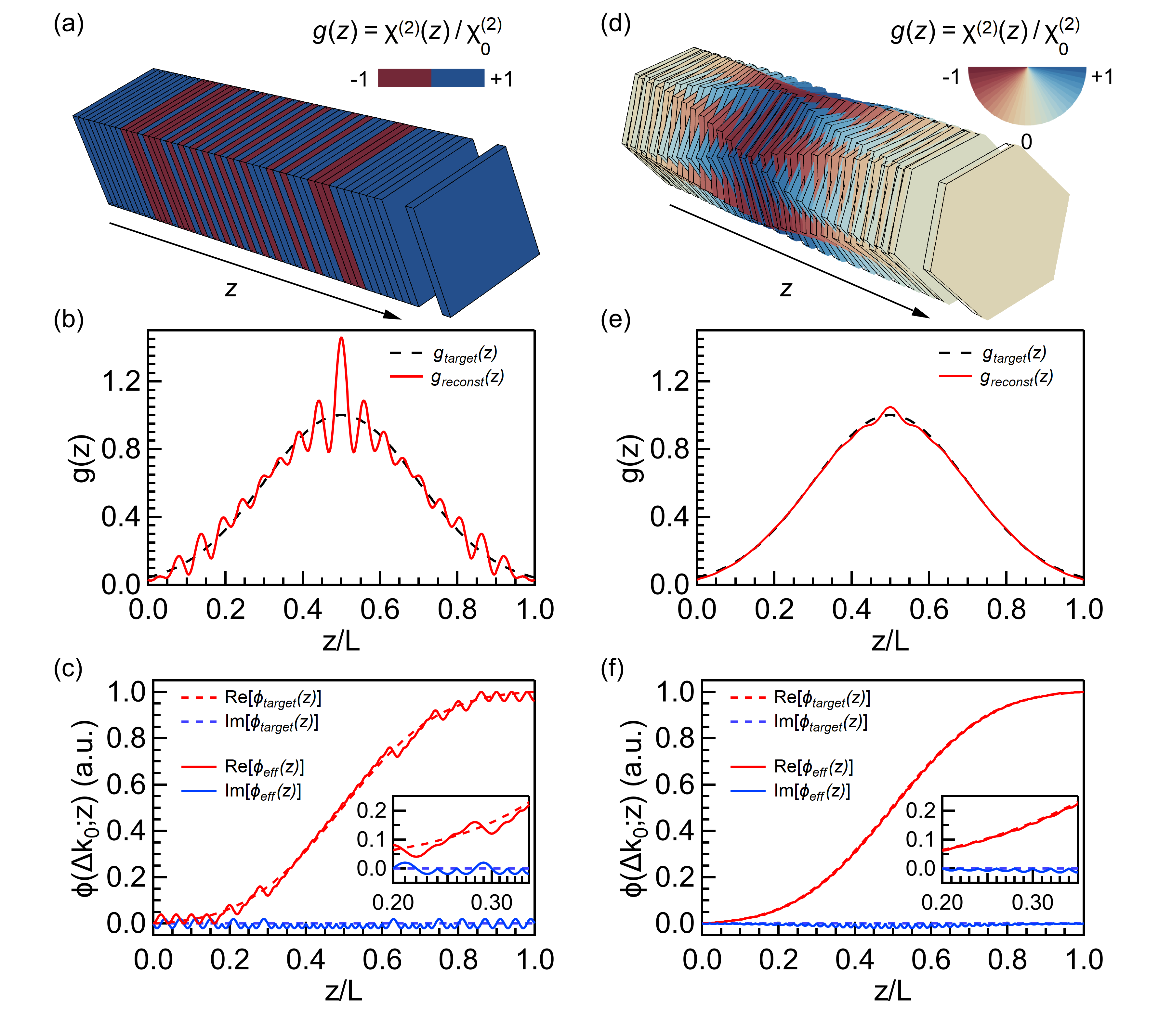}
    \end{center}
    \caption{(a) Schematic of the core concept behind conventional domain-engineering techniques for spectral shaping of SPDC single-photon wave-packets. The relative nonlinear coefficient $\chi^{(2)}_0/|\chi^{(2)}_0|$ is restricted to alternate between +1 and –1 across crystal domains in a non-trivial pattern. (b) Example of a reconstructed normalized nonlinearity profile $g_{\mathrm{reconst}}(z)$ enabled by the non-trivial poling concept depicted in (a), which approximates a target Gaussian function $g_{\mathrm{target}}(z)$. (c) Calculated tracking $\phi_{\mathrm{track}}(z)$ and effective $\phi_{\mathrm{eff}}(z)$ PMFs corresponding to the $g(z)$ functions shown in (b). (d) Schematic of the proposed twist-angle nonlinearity engineering method that enables the design of nonlinear QPM crystals in which $\chi^{(2)}_0/|\chi^{(2)}_0|$ at each crystal domain can be chosen from a discrete set of values between -1 and +1 due to the twist-angle degree of freedom. (e) Example of a reconstructed normalized nonlinearity profile $g_{\mathrm{reconst}}(z)$ enabled by the twist-angle engineering concept depicted in (d), which allows a better approximation to a target Gaussian function $g_{\mathrm{target}}(z)$ than shown in (b). (f) Calculated tracking and effective PMFs corresponding to the $g(z)$ functions shown in (e).}
    \label{fig1}
\end{figure*}

\section{Twist-angle nonlinearity engineering}
The spectral profile of the SPDC photon pairs is governed by the PMF $\phi(\Delta k)$, which encodes the longitudinal nonlinearity profile $g(z)$ together with the momentum mismatch $\Delta k(\omega_s,\omega_i)$ arising from material dispersion. Here, we define the PMF as
\begin{equation}
\label{PMF}
    \phi(\Delta k(\omega_s,\omega_i)) = \int_{-\infty}^{+\infty}g(z)e^{i\Delta kz}dz,
\end{equation}
where $\Delta k(\omega_s,\omega_i) = k_p(\omega_s + \omega_i) - k_s(\omega_s)-k_i(\omega_i)$ represents the momentum mismatch, and `$s$', and `$i$' indicate the two down-converted photons, the `signal' and `idler', respectively. The magnitude $g(z)=\chi^{(2)}(z)/\chi^{(2)}_0$ is the normalized nonlinearity profile at position $z$ in the crystal, which can be tailored via domain engineering. When defined as in Eq. (\ref{PMF}), the PMF exhibits an important property: The PMF for two consecutive crystal domains equals the sum of the PMFs for the individual crystal domains \cite{michler2009quantum,graffitti2017pure}.

Figure \ref{fig1} illustrates the core concept of domain-engineering for spectral shaping of SPDC single-photon wave-packets. In conventional custom-poled nonlinear crystals, $g(z)$ is restricted to alternate between +1 and –1 across crystal domains in a non-trivial pattern [Fig. \ref{fig1}(a)]. This modulation enables shaping of the effective nonlinearity, allowing engineering of the PMF and, consequently, the photon spectral properties \cite{branczyk2010engineered,dixon2013spectral,dosseva2016shaping}. Importantly, as sketched in Fig. \ref{fig1}(b), the non-trivial poling allows the reconstructed nonlinearity profile $g_{\mathrm{reconst}}(z)$ along the QPM crystal to approximate to a target Gaussian function $g_{\mathrm{target}}(z)$ \cite{branczyk2010engineered,dixon2013spectral,dosseva2016shaping,tambasco2016domain,graffitti2017pure}, which is necessary to avoid undesired correlations in the JSA of the two down-converted photons \cite{graffitti2017pure}. Current approaches to custom poling can be broadly classified into two types. The first involves adjusting the widths of the domains within a fixed poling pattern \cite{dixon2013spectral}. The second maintains uniform domain widths, typically equal to or smaller than the coherence length $\ell_c$, but varies the orientation of individual domains instead \cite{branczyk2010engineered,dosseva2016shaping,tambasco2016domain,graffitti2017pure}.

Figure \ref{fig1}(d) shows a sketch of the proposed twist-angle nonlinearity engineering method. In contrast to conventional nonlinear poled crystals, the vdW material platform enables the design of nonlinear crystals in which $g(z)$ at each crystal domain can be chosen from a discrete set of values between -1 and +1. The value of $g(z)$ is determined by the crystal symmetry and the relative twist angle $\theta$ between adjacent domains. In crystals with a three-fold in-plane lattice symmetry, such as transition-metal dichalcogenides (TMDs) and boron nitride (BN), the effective nonlinear coefficient of a pair of twisted layers for co-linearly polarized pump and SPDC photons follows $g(z)=\text{cos}(3\theta)$ \cite{li2013probing,tang2024quasi}. As the layers are rotated, the projection of the nonlinear susceptibility tensor onto the pump and SPDC polarizations varies smoothly, providing a simple and continuous mapping between twist angle and the effective $\chi^{(2)}$. The additional twist-angle degree of freedom offered by the vdW material platform enables finer control over the crystal's nonlinearity profile, allowing for a closer approximation of $g_{\mathrm{reconst}}(z)$ to the optimal Gaussian $g_{\mathrm{target}}(z)$ shape as can be seen in Fig. \ref{fig1}(e).

To showcase the potential of twist-angle nonlinearity engineering in generating spectrally pure down-converted photons, we introduce a novel twist-angle-based domain engineering algorithm. Specifically, we generalize the algorithm in \cite{graffitti2017pure} to account for an arbitrary twist angle $\theta$ between each domain, allowing for a precise approximation of the target crystal nonlinearity profile $g_{\mathrm{target}}(z)$. We begin by summarizing the method introduced in Ref. \cite{graffitti2017pure}. In this domain engineering algorithm, a fixed domain width $w\leq \ell_c$ is first specified, where the coherence length $\ell_c$ is defined in terms of the phase mismatch at the central PDC frequencies $\Delta k(\omega_{s,0},\omega_{i,0}) = \Delta k_0$, $\ell_c=\pi/\Delta k_0$. Although apodized poling patterns with varying domain widths can also be implemented \cite{heese2012role, bostani2015design}, in this work we restrict our analysis to fixed domain widths in order to isolate the role of the twist-angle degree of freedom. The effective nonlinearity profile is then shaped by choosing the orientation of successive domain orientations along the crystal [see Fig. \ref{fig1}(a)]. The decision to flip (or not flip) a given crystal domain is determined by the option that gives the closest approximation to the PMF at the corresponding position $z$ in the crystal. Given a target crystal nonlinearity profile $g_{\mathrm{target}}(z)$ centered at $z=0$, the corresponding PMF at an arbitrary longitudinal position $z$ within the crystal can be calculated as
\begin{equation}
    % \textcolor{red}{A_{target}(z,\Delta k)=\frac{1}{\sqrt{2}}\int^{\frac{L}{2}}_{-\frac{L}{2}}g_{target}(z)e^{-i\Delta k_0z}dz.}
    % \label{fieldamplitude}
    \phi_{\mathrm{track}}(\Delta k;z)=\frac{1}{\sqrt{2}}\int^{z}_{-\frac{L}{2}}g_{\mathrm{target}}(z')e^{-i\Delta kz'}dz',
    \label{fieldamplitude}
\end{equation}
where $g_{\mathrm{target}}(z)$ can be obtained by taking the Fourier transform of the initial target PMF $\phi_{\mathrm{target}}(\Delta k(\omega_s,\omega_i))$ \cite{graffitti2021thesis} (see Supplemental Material for details). Physical intuition for $\phi_{\mathrm{track}}(\Delta k; z)$ arises from its proportionality to the evolution of the signal (idler) field amplitude as it propagates through a crystal with nonlinearity profile $g_{\mathrm{target}}(z)$, assuming a non-depleting pump and a negligible idler (signal) field \cite{graffitti2017pure, tambasco2016domain}. The same reasoning applies to an engineered crystal: the effective phase-matching function $\phi_{\mathrm{eff}}(\Delta k; z)$ is likewise proportional to the field evolution through a crystal characterized by the reconstructed nonlinearity profile $g_{\mathrm{reconst}}(z)$.

Our goal is to design an artificial crystal with a $g_{\mathrm{reconst}}(z)$ corresponding to the target PMF. Since $g(z)$ is independent of $\Delta k$, matching the PMF at a single $\Delta k$ value (usually $\Delta k_0$) is sufficient \cite{graffitti2017pure}. Figures \ref{fig1}(c) and \ref{fig1}(f) show an example of the calculated $\phi_{\mathrm{track}}(\Delta k_0,z)$ and $\phi_{\mathrm{eff}}(\Delta k_0,z)$ corresponding to the $g_{\mathrm{target}}(z)$ and $g_{\mathrm{reconst}}(z)$ shown in Figs. \ref{fig1}(b) and \ref{fig1}(e). Together, Figs.~1(b),(c) and 1(e),(f) illustrate how, as successive domains are added, the cumulative effective PMF $\phi_{\mathrm{eff}}(\Delta k_0,z)$ better tracks the target PMF $\phi_{\mathrm{track}}(\Delta k_0,z)$ when the twist-angle degree of freedom is available, so that the reconstructed nonlinearity profile $g_{\mathrm{reconst}}(z)$ converges toward the desired Gaussian $g_{\mathrm{target}}(z)$.

We now generalize the algorithm presented in Ref. \cite{graffitti2017pure} by incorporating the effects of an arbitrary twist angle between consecutive crystal domains of constant thickness. In a poled crystal fabricated by stacking twisted non-centrosymmetric vdW crystals with in-plane three-fold ($C_3$) rotation symmetry, the PMF at the end of the $m$th domain is given by:
\begin{equation}
    \begin{split}
    \phi_{\mathrm{eff}}&(\Delta k;\{\theta\}_{n=1}^m)=\\&=-i\sum_{n=1}^m\text{cos}(3\theta_n)\int_{(n-1)w}^{nw}e^{i\Delta kz}dz
    \\&=\frac{\bigl(e^{-i\Delta kw}-1\bigl)}{\Delta k}\sum_{n=1}^m\text{cos}(3\theta_n)e^{i\Delta knw},
    \end{split}
\label{cos3theta}
\end{equation}
where $w$ represents the width of the crystal domain and $\theta_n$ represents the twist angle between domains $(n-1)$ and $n$. The domain width $w$ can range from $d\leq w \leq \ell_c$, where $d$ corresponds to the thickness of a single monolayer of the material being simulated (usually $<1$ nm) \cite{laturia2018dielectric}. Note that $\phi_{\mathrm{eff}}(\Delta k;\{\theta\}_{n=1}^m)$ depends on the twist angles of all preceding crystal domains but not on those that follow. It is also worth noting that the method in Ref. \cite{graffitti2017pure} is a special case of our approach, corresponding to the situation where the twist angle is restricted to a binary set of two values differing by $60^{\circ}$ (e.g. only $0^{\circ}$ and $60^{\circ}$). In the following, we will refer to the algorithm as the twist-angle algorithm and the method from \cite{graffitti2017pure} as the sub-$\ell_c$ domains algorithm.

To approximate the $\phi_{\mathrm{track}}$, it is necessary to define a cost function that needs to be minimized at each domain $m$ \cite{graffitti2017pure}:
\begin{equation}
    C_m(\{\theta\}_{n=1}^m)=|\phi_{\mathrm{track}}(\Delta k_0;mw) - \phi_{\mathrm{eff}}(\Delta k_0;\{\theta\}_{n=1}^m)|.
\label{cost}
\end{equation}
A detailed list of the algorithm steps is provided in the Supplemental Material. The procedure can be summarized as follows. First, the domain width $w$, coherence length $\ell_c$, and total number of domains $N$ are defined. Next, the set of allowed twist angles $\theta_{\mathrm{allowed}}$, constrained within  $0^{\circ}\leq\theta_{\mathrm{allowed}}\leq60^{\circ}$ (due to the assumed three-fold $C_3$ symmetry), is specified. The algorithm then proceeds iteratively: starting at $z=0$, the cost function values $C_m(\{\theta_n\}^m_{n=1})$ are computed for the addition of a domain with each possible twist angle in $\theta_{\mathrm{allowed}}$. The domain with the twist angle that yields the lowest cost is selected and added to the structure. This process is repeated until each of the domains has been assigned a twist angle. Overall, this framework establishes twist angle as a practical and flexible design parameter for nonlinear crystal engineering, enabling high-fidelity PMF reconstruction with atomic-scale control.

\section{Benchmark results: idealized \lowercase{vd}W platform}

Before comparing the performance of the different algorithms, we first investigate how the twist-angle step size and domain thickness variables influence the single-photon purities in crystals designed using our twist-angle algorithm. Without loss of generality, our simulations assume a generic non-centrosymmetric vdW crystal with in-plane three-fold $C_3$ rotational symmetry and an ideal symmetric group-velocity matching (GVM) condition at the pump (775 nm) and degenerate signal and idler wavelengths (1550 nm). See Supplemental Material for a discussion of the GVM condition. The single-photon purities are then calculated from the degree of separability of the (JSA) through Schmidt decomposition \cite{laudenbach2016modelling,law2000continuous} (see Supplemental Material for further details). Further details regarding the pump envelop function (PEF), PMF, and the corresponding JSA matrices for the different simulations can also be found in the Supplemental Material. 

\begin{figure}
    \centering
    \includegraphics[width=10.3cm]{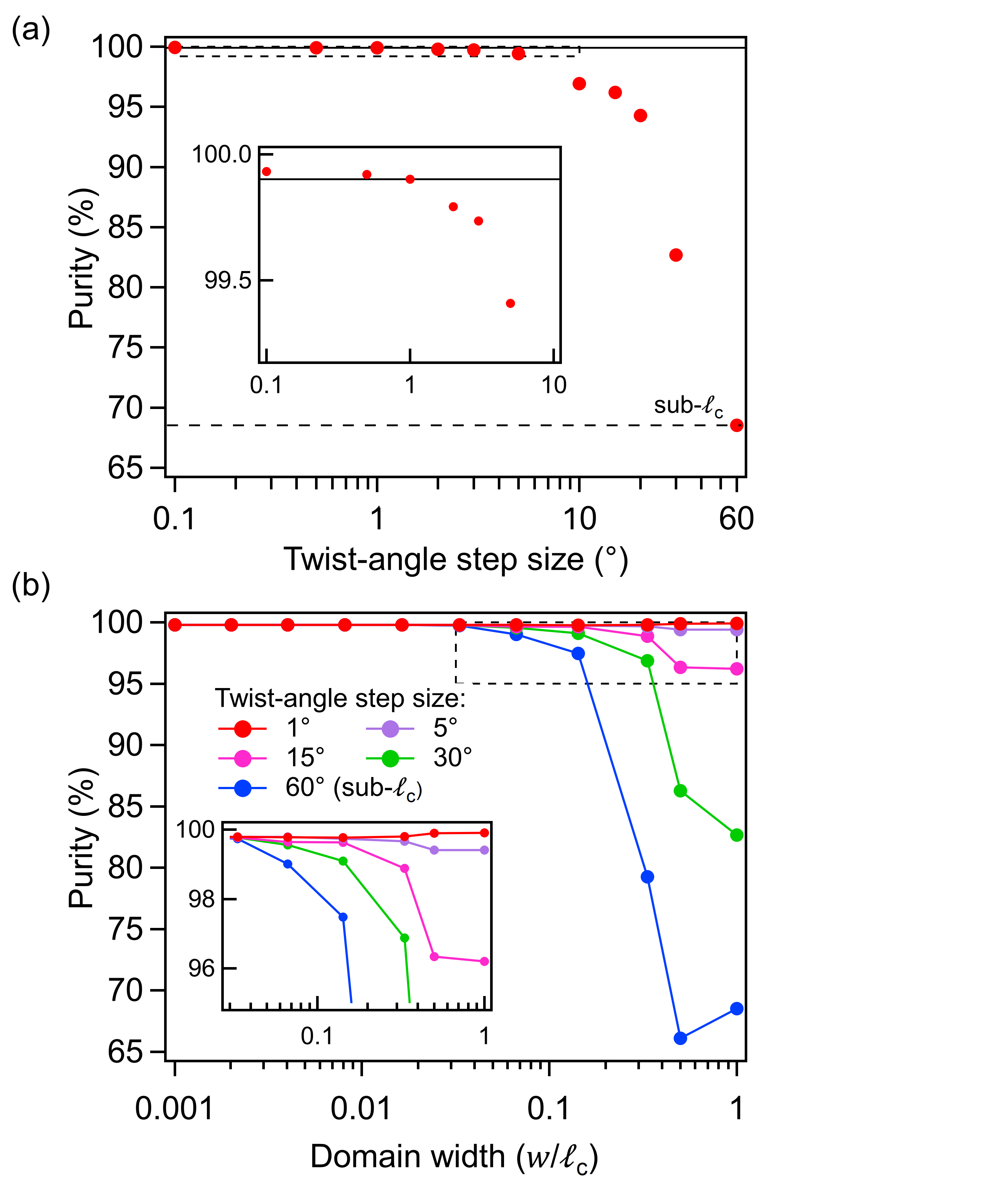}
    \caption{(a) Calculated single-photon purities for twisted nonlinear crystals designed using our twist-angle algorithm, plotted as a function of the step size of the twist angle for a fixed crystal length of 50$\ell_c$. (b) Calculated single-photon purities for twisted nonlinear crystals of total length 50$\ell_c$, designed according to the twist angle algorithm, with different plots representing differing twist-angle step sizes, plotted as a function of the domain width $w$.}
    \label{fig2}
\end{figure}

Figure \ref{fig2}(a) shows the calculated single-photon purities for twisted nonlinear crystals designed using our twist-angle algorithm, plotted as a function of the step size of the twist angle values in $\theta_{\mathrm{allowed}}$. A fixed crystal length of $L=50\ell_c$ and an individual domain width of $w=\ell_c$ are assumed, where $\ell_c$ is used solely as a reference coherence-length scale for normalization (here evaluated using KTP refractive index \cite{zhao2010sellmeier,Konig2004sellmeier}). Consistent with the increased flexibility in shaping the effective nonlinearity afforded by smaller twist-angle steps, the calculated single-photon purity increases as the twist-angle step size is reduced. For these particular crystal parameters, the single-photon purity reaches values $>99.9\%$ for twist-angle steps $\leq1^\circ$ (as indicated by the horizontal black line). The lowest single-photon purity ($\sim68.5\%$, horizontal dashed line) is observed for the crystal with a twist-angle step size of 60$^\circ$, corresponding to a scenario where only two twist angles are allowed ($0^{\circ}$ and $60^{\circ}$). In this limit, the twist-angle algorithm reduces to a binary nonlinearity modulation analogous to conventional sub-$\ell_c$ domain-engineering approaches.

Figure \ref{fig2}(b) shows the calculated single-photon purities for a twisted nonlinear crystal of total length $L=50\ell_c$, plotted as a function of the domain width $w$, in the range $0.001\ell_c \leq w \leq \ell_c$. The different colored dots represent purities calculated for crystals engineered using the twist angle algorithm with varying twist-angle step size. Similarly to the effect of the twist-angle step size, finer shaping of the nonlinearity—achieved through narrower domain widths—leads to higher single-photon purities, with the improvement being more pronounced in crystals designed with larger twist-angle step sizes. Interestingly, for crystals designed with a $1^{\circ}$ twist-angle step, the calculated single-photon purity remains almost constant at approximately $99.9\%$, independent of the domain width. The near-flat dependence of the purity on domain width in this regime shows that twist-engineered vdW crystals can operate with experimentally convenient domain sizes up to $w \approx \ell_{\mathrm{c}}$ without sacrificing performance, alleviating the need for extremely fine or densely patterned domains over extended interaction lengths.

\begin{figure}
    \centering
    \includegraphics[width=8.7cm]{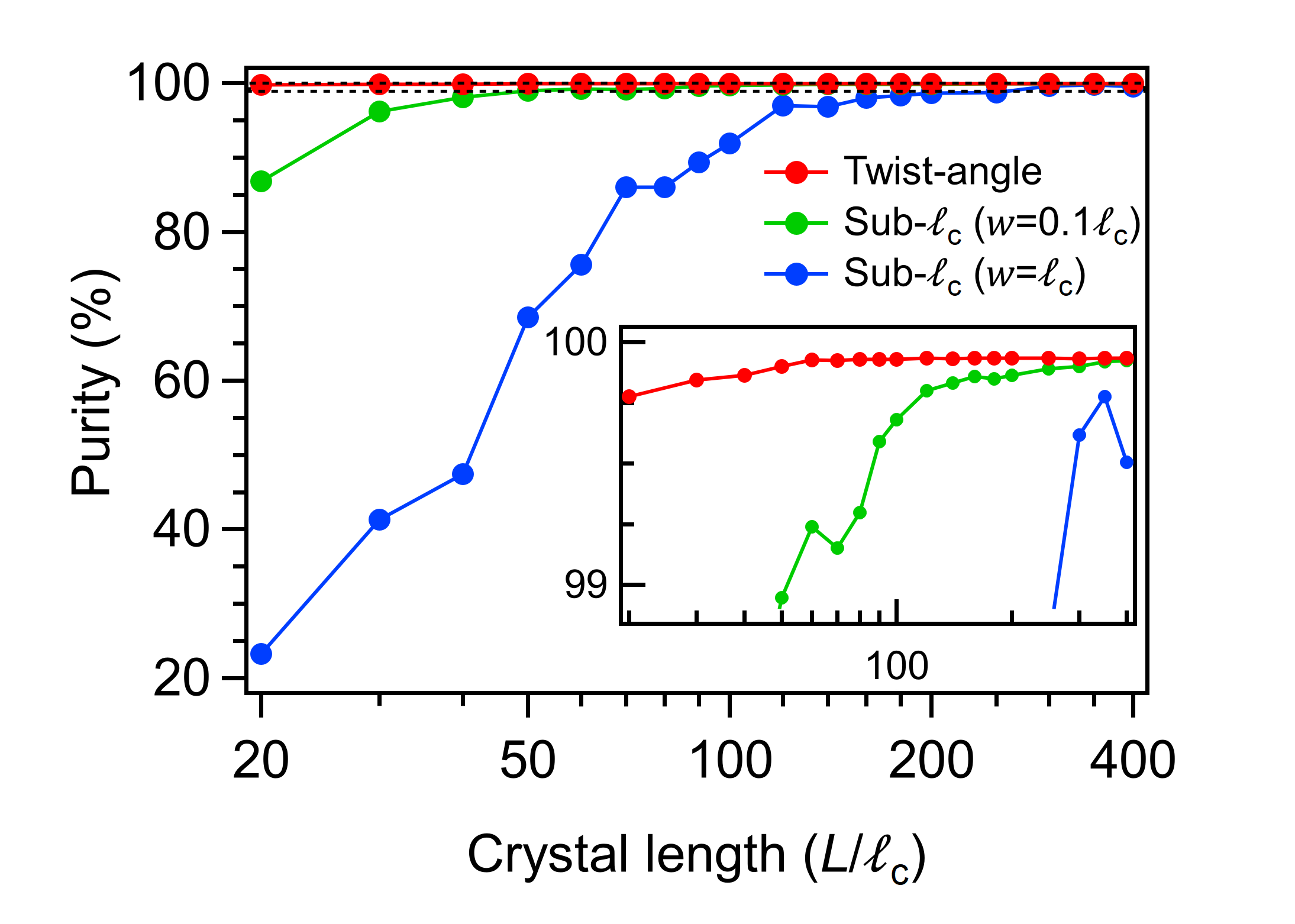}
    \caption{Calculated single-photon purity as a function of total crystal length $L$ for crystals engineered using different nonlinearity-shaping algorithms. Red dots correspond to crystals designed using the twist-angle domain-engineering algorithm with a fixed twist-angle step size of $1^\circ$ and a fixed domain width of $w=\ell_c$. Green and blue dots show results for the sub-$\ell_c$ domains algorithm with fixed domain widths of $w=0.1\ell_c$ and $w=\ell_c$, respectively.}
    \label{fig3}
\end{figure}

Next, we investigate the performance of crystals engineered using different algorithms as a function of the total crystal length $L$, while fixing the domain width in each case. Figure~\ref{fig3} shows the calculated single-photon purities for crystals with lengths ranging from $20\ell_c$ to $400\ell_c$, designed using both the twist-angle domain-engineering algorithm and the sub-$\ell_c$ domains algorithm.

For the twist-angle algorithm, we consider crystals engineered with a fixed twist-angle step size of $1^\circ$ and a fixed domain width of $w=\ell_c$ (red dots). This choice is motivated by two considerations: (i) our results in Fig.~\ref{fig2}(b) show that, for a $1^\circ$ twist-angle step, the single-photon purity is largely insensitive to the domain width; and (ii) a $1^\circ$ twist-angle step is readily achievable using current vdW layer assembly techniques (see Discussion).

For comparison, the green and blue dots show results for crystals engineered using the sub-$\ell_c$ domains algorithm with fixed domain widths of $w=0.1\ell_c$ and $w=\ell_c$, respectively. Across the full range of crystal lengths considered ($20 \leq L/\ell_c \leq 400$), crystals engineered with the twist-angle algorithm consistently yield single-photon purities exceeding $99.7\%$. In particular, for shorter crystals with $L \lesssim 50\ell_c$, the twist-angle algorithm substantially outperforms the sub-$\ell_c$ domains approach, highlighting its advantage for achieving high single-photon purity in shorter interaction lengths.

As the crystal length increases, the differences in single-photon purity between the various designs diminish, with all approaches converging to purities above $99.7\%$ for $L=400\ell_c$. These results are especially relevant for the vdW platform, where precise control over interlayer twist angle currently exceeds the ability to assemble long multilayer stacks with uniformly high fidelity (see Discussion). Examples of the optimized twist-angle profiles for representative crystal lengths are provided in the Supplemental Material.

Taken together, these results demonstrate that continuous control of $\chi^{(2)}$ via interlayer twist enables high-fidelity reconstruction of target phase-matching functions across a broad range of device lengths and domain-engineering parameters, establishing twist-engineered vdW crystals as a practical route to compact, high-purity SPDC sources under idealized conditions.

\section{Twist-engineered SPDC in highly non-degenerate wavelength regimes}

Beyond the degenerate, telecom-centric regime considered in an arbitrary crystal above, we now examine the performance of twist-angle domain engineering in realistic, non-centrosymmetric vdW crystals and in strongly non-degenerate wavelength configurations that are challenging for conventional bulk nonlinear platforms \cite{boyd2008nonlinear, kuo2006optical}.

\begin{figure}
    \centering
    \includegraphics[width=9cm]{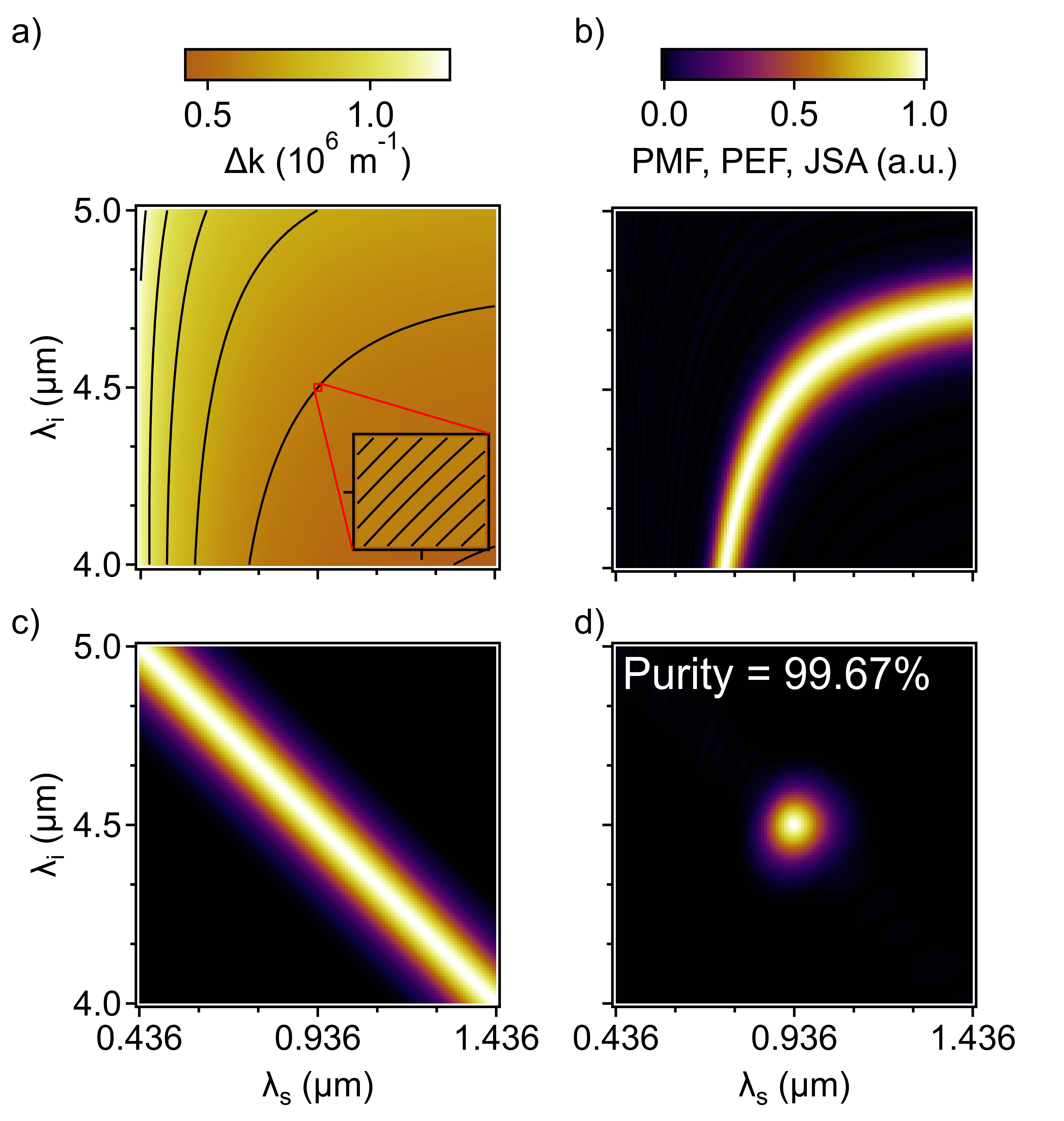}
    \caption{Calculated joint-spectral quantities used to evaluate SPDC photon purity for a non-degenerate wavelength configuration in rBN. (a) Calculated momentum mismatch $\Delta k(\lambda_s,\lambda_i)$; the inset highlights the approximately linear behavior of momentum mismatch contours over a narrow wavelength range. (b) PMF obtained via twist-angle domain engineering for the configuration considered. (c) Corresponding pump envelope function (PEF), which together with the PMF determines the spectral purity of the generated photon pairs. (d) Resulting JSA, exhibiting a near-factorable structure and yielding a calculated single-photon purity of $99.67\%$ without spectral filtering.
}
    \label{fig4}
\end{figure}

As a representative example, we consider a twist-engineered rBN crystal, using experimentally reported refractive index data for hexagonal boron nitride (hBN) to model the material dispersion \cite{segura2018natural}. We focus on a highly asymmetric wavelength configuration in which the idler photon lies in the mid-infrared at $\lambda_i = 4.5~\mu$m, while the signal photon is in the near-infrared around $\lambda_s \approx 936$~nm, corresponding to a pump wavelength of 775~nm. Figure~\ref{fig4} summarizes the resulting phase-matching and spectral properties of the twist-engineered crystal. Figure~\ref{fig4}(a) shows the calculated momentum mismatch $\Delta k(\lambda_s,\lambda_i)$ for rBN. Two features are particularly relevant: (i) the phase-matching contours exhibit slopes between $0^{\circ}$ and $90^{\circ}$, and (ii) as highlighted in the inset, the contours are approximately linear over small wavelength intervals. These characteristics are consistent with conditions favorable for group-velocity matching \cite{graffitti2018design, graffitti2021thesis} and suggest that high-purity photon generation is feasible in rBN even in wavelength regimes that are challenging to access using conventional bulk nonlinear crystals. Further discussion of the role of these features in enabling GVM is provided in the Supplemental Material.

Combining this $\Delta k(\lambda_s,\lambda_i)$ landscape with twist-angle domain engineering using parameters $L = 50\ell_c$, $w = 1\ell_c=5.28~\mu$m, and a twist-angle step size of $1^{\circ}$ yields an optimized PMF, shown as a function of $(\lambda_s,\lambda_i)$ in Fig.~\ref{fig4}(b), together with the corresponding width-optimized PEF in Fig.~\ref{fig4}(c). The resulting JSA, shown in Fig.~\ref{fig4}(d), exhibits a near-factorable structure and yields a single-photon spectral purity of $99.67\%$.

To assess the general applicability of pure photon generation in real twist-engineered crystals, analogous calculations for other experimentally relevant non-centrosymmetric vdW crystals have been performed, including 3R-WS$_2$ and 3R-MoS$_2$, using reported material parameters \cite{haastrup2018the, gjerding2021recent}. In both cases, twist-angle domain engineering enables effective shaping of the PMF and yields near-factorable JSAs with high single-photon purity, demonstrating that the approach extends beyond rBN to a broader class of vdW materials. Additional details and results for these materials are provided in the Supplemental Material.

These results demonstrate that the twist-angle domain-engineering framework is applicable to real vdW crystals and remains effective in regimes characterized by large signal--idler wavelength separation and strong dispersion asymmetry. Importantly, this performance is achieved using realistic material parameters, underscoring the robustness of continuous $\chi^{(2)}$ engineering via interlayer twist beyond idealized scenarios.

\section{Discussion and outlook}

It is important to emphasize that the generation of pure photons is only one of many applications of domain engineering. Chirped gratings \cite{carrasco2004enhancing}, pulse-mode engineering \cite{graffitti2020direct}, and frequency-bin entanglement \cite{morrison2022frequency} are all examples of domain-engineering applications well-suited to harness the unique properties of vdW materials. In particular, twist-angle nonlinearity engineering excels in applications that require the faithful reconstruction of complex phase-matching functions or nonlinearity profiles. The Supplemental Material includes a brief showcase demonstrating how twist-engineered vdW crystals enable high-fidelity approximations to nonlinearity profiles beyond the Gaussian, such as first- and second-order Hermite–Gauss modes. 

Beyond these idealized benchmark scenarios, our results demonstrate that twist-angle domain engineering remains effective in realistic, non-centrosymmetric vdW crystals and in strongly non-degenerate wavelength regimes. Using rBN as a representative example, we show that high-purity SPDC can be achieved for highly asymmetric signal--idler wavelength configurations spanning the near- and mid-infrared. To assess the generality of this behavior, we have also performed analogous calculations for other experimentally relevant vdW materials, including twist-engineered 3R-WS$_2$ and 3R-MoS$_2$. In all cases considered, continuous control of $\chi^{(2)}$ via interlayer twist enables effective shaping of the PMF and yields near-factorable JSAs with high single-photon purity, underscoring the robustness of the proposed framework beyond a single material platform. While photon purity is used here as a benchmark, the same approach can in principle be extended to tailor other properties of non-degenerate SPDC sources, including bandwidth asymmetry, frequency separation, and temporal mode structure.

We now comment on the experimental feasibility of implementing the twist-angle algorithm for the design of tailored nonlinear vdW crystals. One important consideration is that the $g(z)=\cos(3\theta)$ dependence used in our calculations corresponds to one experimental configuration in which the pump and down-converted photons are co-linearly polarized. This condition arises because the linear polarization of the generated SPDC photons depends on the relative orientation between the pump polarization and the armchair crystal axis of the domain in which they are generated \cite{trovatello2021optical, hong2023twist, weissflog2024tunable}. Twist-angle domain engineering with circularly polarized pump, signal, and idler photons is also feasible, but lies beyond the scope of the present work.

Another key experimental consideration in designing twisted nonlinear crystals is the achievable twist-angle step size. Current van der Waals assembly techniques routinely attain twist-angle steps well below $1^\circ$ \cite{cao2018unconventional,liao2020precise,kwanghee2025highly}. This level of control enables the realization of crystals with high photon purities even for domain widths approaching $\ell_c$, thereby reducing the number of layers that must be stacked and simplifying fabrication. Moreover, recent progress in vdW assembly—most notably through autonomous robotic stacking—has demonstrated artificial crystals incorporating up to 80 layers \cite{masubuchi2018autonomous,mannix2022robotic}, suggesting that such structural control may soon be routinely achievable. Furthermore, even in crystals designed with a twist angle step size of $1^\circ$, the relative twist angle between adjacent domains generally exceeds $\sim$$2^\circ$–$3^\circ$ (see Supplemental Material). This implies that atomic lattice reconstruction is unlikely to occur at domain interfaces \cite{weston2020atomic,enaldiev2020stacking,halbertal2021moire}.

In summary, we have introduced a general algorithm that leverages the twist-angle degree of freedom unique to the vdW material platform to design artificial crystals with tailored longitudinal nonlinearity profiles. Using SPDC as a benchmark, we demonstrate that twist-angle domain engineering enables high-fidelity reconstruction of target phase-matching functions and significantly enhances single-photon purity compared to conventional domain-engineering approaches, particularly in compact devices. We further show that this framework remains effective across multiple experimentally relevant vdW materials and in strongly non-degenerate wavelength regimes, underscoring its robustness beyond near-degenerate and idealized scenarios. Together, these results position vdW heterostructures as a versatile platform for engineered nonlinear-optical functionalities, combining atomically precise $\chi^{(2)}$ control with scalable fabrication pathways for next-generation quantum light sources.

\textit{Acknowledgments} -  {We thank Jonathan Leach for fruitful discussions. This work was supported by the EPSRC (grant no. EP/Z533208/1). D. M. is supported by an EPSRC Doctoral Landscape Award. M.B.-G. is supported by a Royal Society University Research Fellowship. B.D.G. is supported by a Chair in Emerging Technology from the Royal Academy of Engineering.

\bibliography{combined_bib}
\newpage
\textcolor{white}{...}
\newpage
\onecolumngrid
\clearpage
\renewcommand{\thefigure}{S1}
\renewcommand{\thetable}{S1}

\section*{Supplementary Material 1: Domain engineering theory}

\noindent Figure \ref{figs1} illustrates the concept of nonlinearity engineering. To achieve maximal single-photon spectral purity, a Gaussian PMF is required \cite{nicolas2018gaussian}. The target Gaussian PMF, $\phi_{\mathrm{target}}(\Delta k)$, shown in Fig. \ref{figs1}(a), is defined as
\begin{equation}
\tag{S1}
    \phi_{\mathrm{target}}(\Delta k)=e^{-\frac{1}{2}(\Delta k - \Delta k_{0})^{2}{\sigma^2}},
    \label{pmf}
\end{equation}
where $\sigma$ denotes the width of the PMF. The Fourier transform of $\phi_{\mathrm{target}}(\Delta k)$ yields the corresponding target nonlinearity profile $g_{\mathrm{target}}(z)$, shown in Fig. \ref{figs1}(b):
\begin{equation}
\tag{S2}
    \label{gtarget}
    g_{\mathrm{target}}(z)=\frac{1}{\sigma} e^{-\frac{z^{2}}{2{\sigma^{2}}}+i\Delta k_{0}z},
\end{equation}
where it can be seen that both $\phi_{\mathrm{target}}(\Delta k)$ and $g_{\mathrm{target}}(z)$ have a dependence on $\sigma$. Note that this equation results in a crystal centred at $z=0$. 

\begin{figure}[h!]
    \centering
    \includegraphics[width=13cm]{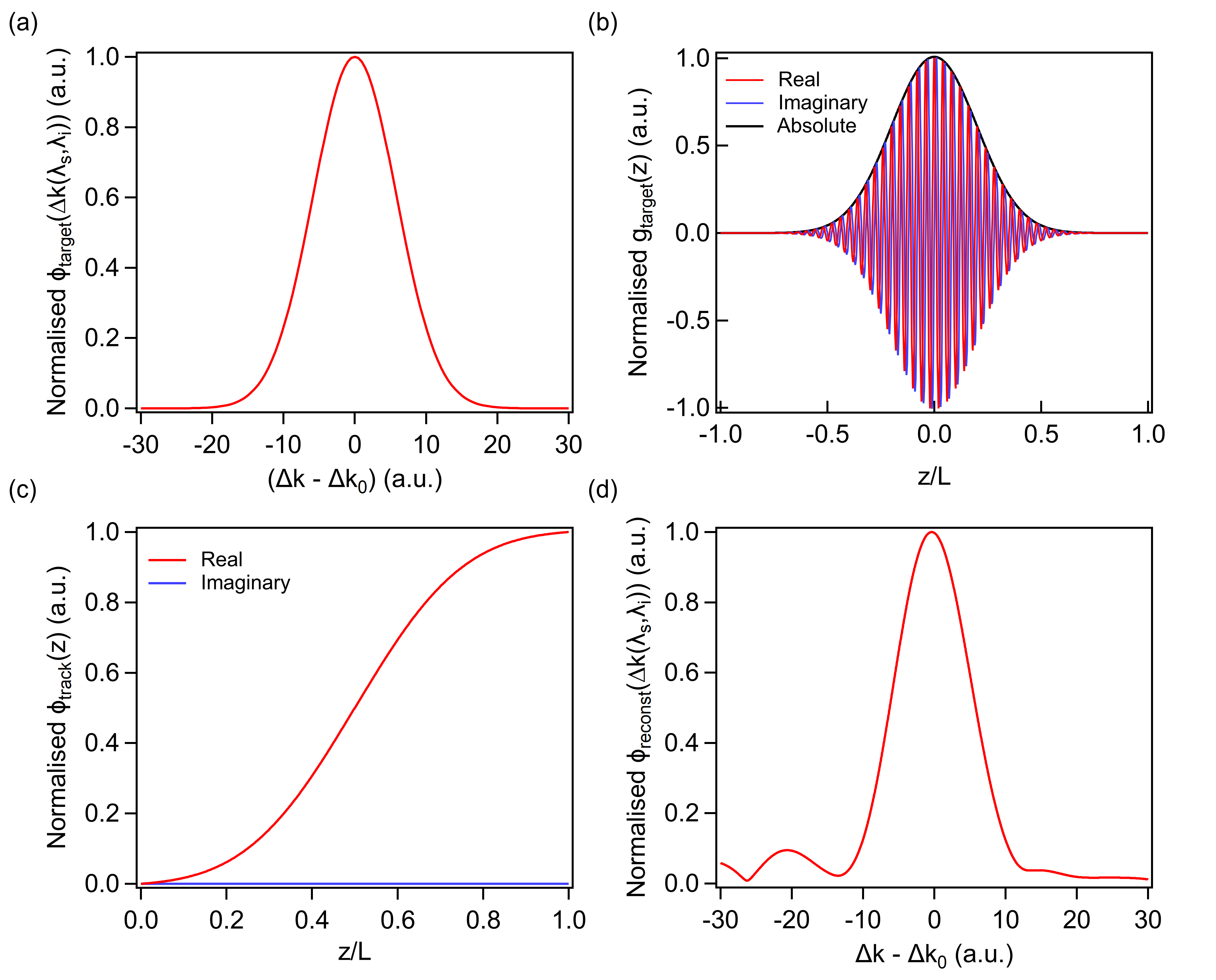}
    \caption{(a) Normalised target PMF $\phi_{\mathrm{target}}(\Delta k)$, plotted with the x-axis centred at a chosen momentum mismatch value. (b) Normalised target nonlinearity profile $g_{\mathrm{target}}(z)$, obtained as the Fourier transform of (a), plotted as a function of the normalised crystal length $L$ (with $L = 0$ at the crystal centre). (c) Normalised PMF for $\Delta k = \Delta k_0$ $\phi_{\mathrm{track}}(z)$, plotted as a function of position inside the crystal, corresponding to the nonlinearity profile in (b). (d) Reconstructed PMF $\phi_{\mathrm{reconst}}(\Delta k)$, calculated by applying a domain-engineering algorithm to approximate $\phi_{\mathrm{track}}(z)$. Note that $\phi_{\mathrm{reconst}}(\Delta k)$ in (d) is less smooth than $\phi_{\mathrm{target}}(\Delta k)$ in (a), since the chosen algorithm input parameters limit the approximation quality.}
    \label{figs1}
\end{figure}

When choosing $\sigma$, a trade-off must be made: smaller values yield higher spectral purity but reduce system efficiency \cite{tambasco2016domain}. In this work, we set $\sigma = L/5$, where $L$ is the crystal length, following previous recommendations in the literature as a suitable compromise \cite{graffitti2021thesis}. An equation for the PMF which can be tracked longitudinally through the crystal is now required. To find this, we perform an inverse Fourier transform of $g_{\mathrm{target}}(z)$, with integration boundaries corresponding to the edges of the crystal as defined in Eq. (\ref{gtarget}). This procedure yields a position-dependent equation for the PMF \cite{graffitti2021thesis, tambasco2016domain}:

\begin{equation}
\tag{S3}
    \begin{split}
    \phi_{\mathrm{track}}(\Delta k_0;z)&=\frac{1}{\sqrt{2}}\int^{z}_{-\frac{L}{2}}g_{\mathrm{target}}(z')e^{-i\Delta k_0z'}dz'\\&=\sqrt{\frac{2}{\pi}}\sigma\left(\mathrm{erf}\left[\frac{L}{2\sqrt{2}\sigma}\right]+\mathrm{erf}\left[\frac{z-\frac{L}{2}}{\sqrt{2}\sigma}\right]\right).
    \end{split}
\label{fieldamplitude}
\end{equation}

Given a crystal with in-plane three-fold $C_3$ rotational symmetry in which an arbitrary twist is applied to each domain, the corresponding PMF at each point can be calculated using Eq. (3) from the main text. To engineer a crystal that heralds photons with a target PMF, $\phi_{\mathrm{target}}(\Delta k)$, we reverse the above procedure by determining the domain orientations that best reproduce $\phi_{\mathrm{track}}(\Delta k;z)$. The optimal domain configuration is obtained using a domain-engineering algorithm.

\section*{Supplementary Material 2: Algorithm steps}

\noindent Here, we provide a detailed walk-through of the steps in the twist-angle algorithm. This algorithm generalizes the arbitrarily-small domain method from \cite{graffitti2017pure}, with the two being equivalent when $\theta_{\mathrm{allowed}}={0^{\circ}, 60^{\circ}}$. Note also that the implementation uses list indexing starting at 0.

\begin{enumerate}
  \item Compute $\phi_{\mathrm{track}}(z)$.
  \item Define the domain width $w$, number of domains $N$, coherence length $\ell_c$, and list of permitted twist-angles $\theta_{\mathrm{allowed}}$.
  \item Initialise a list $S$ of length $N$ to store domain orientations, setting the first orientation to $0^{\circ}$.
  \item Set $n = 0$.
  \item Generate possible orientation lists of length $n$ by appending a domain with each $\theta_{\mathrm{allowed}}$ value to the list S.
  \item Evaluate the cost function for each candidate orientation list.
  \item Assign $S[n]$ as the orientation yielding the lowest cost.
  \item Update $n \mapsto n+1$ and repeat steps 5–7 until all $N$ domains are assigned an orientation.
\end{enumerate}

Figure~\ref{twist_angles} shows examples of the optimized twist-angle distributions obtained using the twist-angle algorithm for engineered twisted crystals comprising $N=50$ and $N=20$ domains. In both cases, the domains have a width $w=\ell_c$, and a twist-angle step size of $1^{\circ}$ is used (i.e., $\theta_{\mathrm{allowed}}={0^{\circ},1^{\circ},2^{\circ},\dots,60^{\circ}}$).

\renewcommand{\thefigure}{S2}
\begin{figure}[H]
    \centering
    \includegraphics[width=13cm]{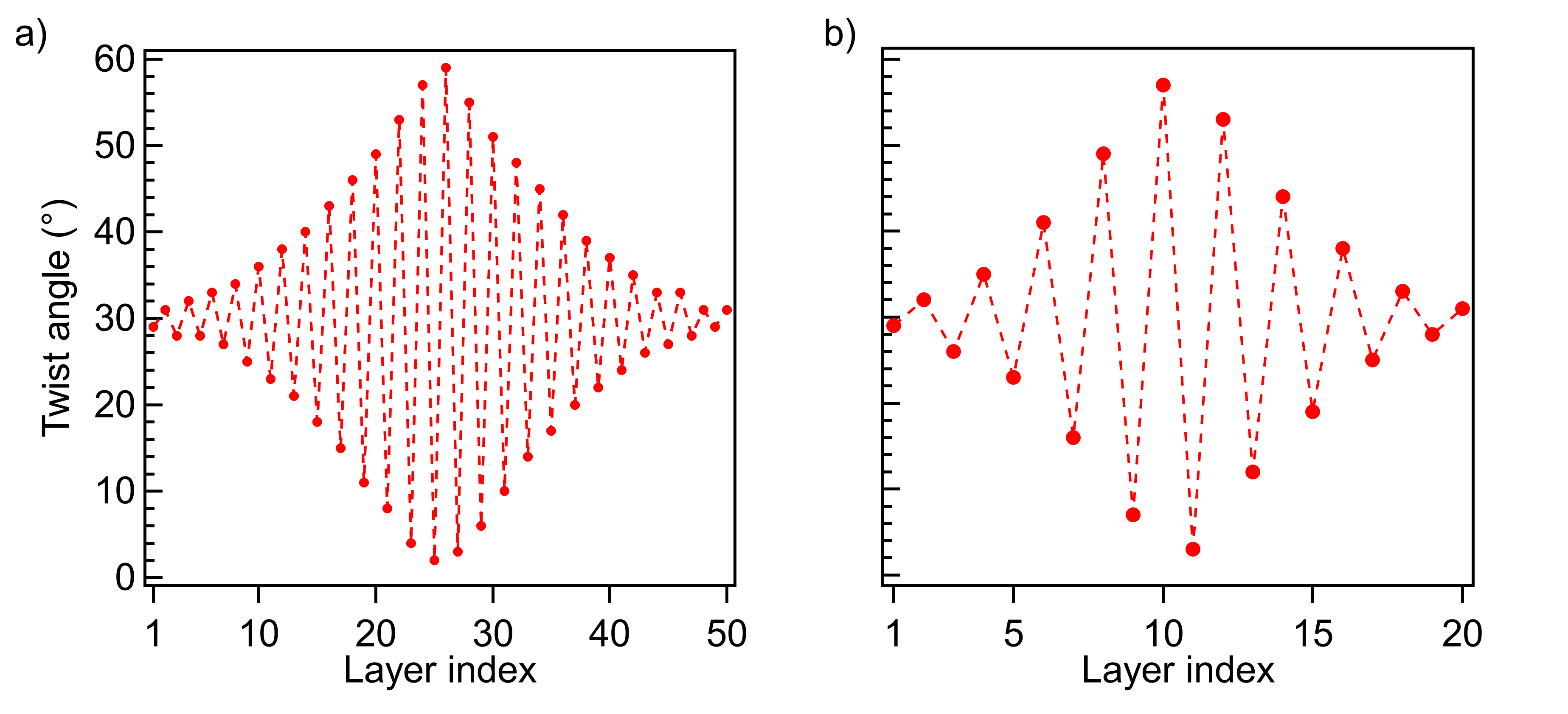}
    \caption{Twist angles calculated using the twist-angle algorithm. (a) Using input parameters $w=\ell_c$, $N=50$, and $\theta_{\mathrm{allowed}}={0^{\circ},1^{\circ},2^{\circ},\dots,60^{\circ}}$. (b) Identical input parameters as (a) apart from $N=20$. Angles are plotted relative to the orientation of the first domain.}
    \label{twist_angles}
\end{figure}

\section*{Supplementary Material 3: Group-velocity matching condition}

Group-velocity matching (GVM) is a widely used approach for reducing spectral correlations between signal and idler photons. For a chosen set of pump, signal, and idler wavelengths, suitable phase-matching conditions can be identified—together with an optimal pump bandwidth and crystal length—that align the group velocities of the pump and down-converted photons. This suppresses correlations arising from relative photon arrival-time differences at the detector \cite{mosley2008heralded,graffitti2017pure}. Under these conditions, a Gaussian pump envelope combined with a Gaussian PMF can yield a highly separable JSA and, consequently, high photon purity.

\renewcommand{\thefigure}{S3}
\begin{figure}[h]
    \centering
    \includegraphics[width=13cm]{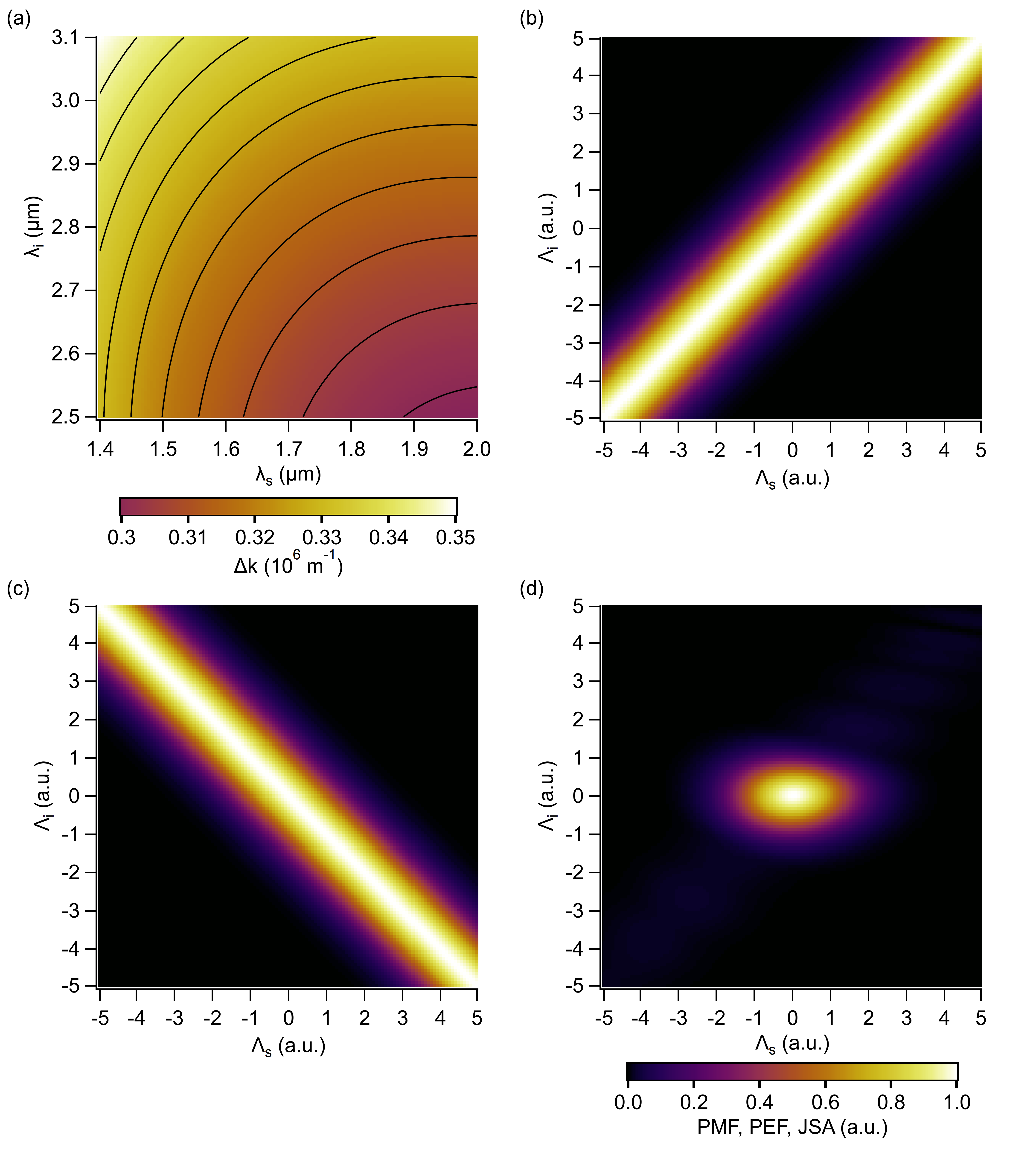}
    \caption{(a) An example of momentum mismatch $\Delta k$ for the 2D crystal hBN as a function of the signal and idler photon wavelengths. A wavelength range where the $\Delta k$ contours are both angled between $0^{\circ}$ and $90^{\circ}$ and close to linear needs to be chosen to fulfill the GVM condition. (b) An ideal Gaussian PMF, plotted as function of the shifted signal and idler wavelengths and using a linear momentum mismatch. (c) An example Gaussian PEF, which indicates signal-idler wavelength pairs that are possible for a set of pump properties. Note that the PEF will always run from the top left to bottom right in the $(\Lambda_i,\Lambda_s)$ space due to conservation of energy. (d) An example JSA calculated as the product of the PMF and PEF shown in panels (b) and (c), respectively. The displayed JSA would result in the generation of photons with $100\%$ purity, as the PEF and PMF are both perfectly Gaussian and obey the GVM condition.}
    \label{PMF_PEF_JSA}
\end{figure}

Figure~\ref{PMF_PEF_JSA}(a) shows the calculated momentum mismatch $\Delta k(\lambda_s,\lambda_i)$ for the vdW crystal rBN, evaluated using the dispersion relation of hBN \cite{segura2018natural}, as a function of the signal and idler wavelengths. Because the PMF depends on $\Delta k$, a PMF plotted for this material would follow the contours of $\Delta k$ in the same parameter space. Achieving high photon purity requires selecting central signal and idler wavelengths for which $\Delta k$ varies approximately linearly \cite{graffitti2021thesis}. Over a narrower wavelength range than that shown in Fig.~\ref{PMF_PEF_JSA}(a), the hBN contours exhibit near-linear behavior, indicating that group-velocity matching can be achieved for specific wavelength configurations—an essential condition for the practical implementation of twist-angle nonlinearity engineering in vdW crystals.

The ideal Gaussian PMF shown in Fig.~\ref{PMF_PEF_JSA}(b) is calculated for an arbitrary material exhibiting a perfectly linear momentum mismatch. In this idealized case, the contours of $\Delta k$, plotted as a function of the signal and idler wavelengths, run from the bottom left to the top right at a $45^{\circ}$ angle with no curvature. Linear momentum mismatch is consistently assumed throughout this work when benchmarking and comparing different domain-engineering algorithms, ensuring that the resulting comparisons are independent of specific material dispersion properties. Figure~\ref{PMF_PEF_JSA}(b) is expressed in terms of $\Lambda_s$ and $\Lambda_i$, which denote the signal and idler wavelengths shifted relative to their chosen central values.

The second function required to construct the JSA is the pump-envelope function (PEF), which encodes the properties of the pump field driving the down-conversion process, including its central wavelength, spectral shape, and bandwidth. For a Gaussian pump, the PEF can be written as:
\begin{equation}
\tag{S4}    \alpha(\lambda_s,\lambda_i)=\mathrm{exp}\left({-\frac{(\Lambda_s+\Lambda_i)^2}{2\sigma^2_{PEF}}}\right),
    \label{PEF}
\end{equation}
where $\sigma_{\mathrm{PEF}}$ denotes the standard deviation of the Gaussian PEF. Throughout this work, a Gaussian PEF is used, as it maximizes photon purity when paired with a Gaussian PMF \cite{graffitti2018design}. An example of the PEF defined above is shown in Fig.~\ref{PMF_PEF_JSA}(c), plotted in terms of $\Lambda_s$ and $\Lambda_i$. Owing to energy conservation, the PEF is oriented along the top-left to bottom-right diagonal in this parameter space. When both the PMF and PEF are Gaussian and defined on the same axes, the group-velocity-matching (GVM) condition can be expressed as \cite{ansari2018tailoring}:
\begin{equation}
\tag{S5}
    2 \text{cos}(\theta)\text{sin}(\theta) = \frac{\sigma^2_{PMF}}{\sigma^2_{PEF}}.
    \label{perfectGVM}
\end{equation}
Here, $\sigma_{\mathrm{PMF}}$ denotes the standard deviation of the PMF when expressed in the $(\Lambda_s,\Lambda_i)$ parameter space (not to be confused with the previously defined $\sigma$), and $\theta$ is the angle of the PMF relative to the $x$-axis. The analytical relation between $\sigma_{\mathrm{PMF}}$ and $\sigma_{\mathrm{PEF}}$ given in Eq.~\ref{perfectGVM} can deviate slightly from the true optimum when realistic crystal dispersion is considered, rather than an idealized perfectly linear $\Delta k(\lambda_s,\lambda_i)$. Consequently, when working with real material parameters, it is beneficial to explore $\sigma_{\mathrm{PEF}}$ values in the vicinity of the analytical prediction. The GVM condition requires $0^{\circ} < \theta < 90^{\circ}$, with
\begin{equation}
\tag{S6}
    \text{tan}(\theta)=-\frac{v_{p}^{-1}-v_{s}^{-1}}{v_{p}^{-1}-v_{i}^{-1}},
\end{equation}
where $v_{p,s,i}$ are the group velocities of the pump, signal, and idler photons, respectively \cite{graffitti2018design, ansari2018tailoring}. The joint spectral amplitude (JSA) is then defined as the product of the PEF and PMF:
\begin{equation}
\tag{S7}
    f(\lambda_s,\lambda_i) = \alpha(\lambda_s,\lambda_i)\phi(\Delta k(\lambda_s, \lambda_i)).
    \label{JSA}
\end{equation}

Figure~\ref{PMF_PEF_JSA}(d) shows an example JSA obtained from the PMF and PEF in Figs.~\ref{PMF_PEF_JSA}(b) and \ref{PMF_PEF_JSA}(c), respectively. Once the JSA is computed, the photon spectral purity can be evaluated using the Schmidt decomposition \cite{graffitti2018design}.

\section*{Supplementary Material 4: Schmidt decomposition}

Spectral correlations between the signal and idler photons, and thus the photon purity, can be quantified via the degree of separability of the JSA \cite{graffitti2017pure, graffitti2018design}, as determined using Schmidt decomposition. The JSA is first expressed as a sum of orthogonal modes:
\begin{equation}
\tag{S8}
f(\omega_s,\omega_i) = \sum_k b_k u_k(\omega_s) v_k(\omega_i),
\end{equation}
where $u_k(\omega_s)$ and $v_k(\omega_i)$ denote the Schmidt modes and $b_k$ the corresponding Schmidt coefficients. The Schmidt modes represent orthogonal single-photon spectral distributions, while the Schmidt coefficients are real numbers whose squared sum equals unity for a normalized JSA $f(\omega_s,\omega_i)$. The JSA purity is then given by
\begin{equation}
\tag{S9}
\mathcal{P} = \sum_k b_k^4.
\label{schmidtsum}
\end{equation}
Since an analytical determination of the Schmidt coefficients is generally not possible, a singular value decomposition (SVD) of the numerically constructed JSA is used instead. The singular values obtained from the SVD replace the Schmidt coefficients in Eq. \ref{schmidtsum}.

\indent When constructing the JSA numerically, appropriate choices for the spectral range $\zeta$ and the number of discretized frequency points $D$ are critical, as both parameters influence the calculated photon purity. Here, $\zeta$ is defined as the ratio between the total spectral range of the JSA matrix and its full width at half maximum (FWHM), where the JSA FWHM is defined as $(\lambda_{s\mathrm{FWHM}}+ \lambda_{i\mathrm{FWHM}})/2$. 

\indent In this work, we use $D=125$ and $\zeta \approx 10$, consistent with literature recommendations for accurate purity calculation \cite{graffitti2018design}. While larger values of $\zeta$ would in principle be desirable, the choice $\zeta \approx 10$ is motivated by the short lengths of some simulated crystals (down to $L = 20\ell_c$). Because the JSA bandwidth increases as the crystal length decreases, using $\zeta > 10$ in these cases would extend the spectral grid into unphysical negative wavelength regions, causing the SVD procedure to fail.

\indent In one instance, specifically when calculating the heralded spectral purity for twist-engineered rBN, the relatively shallow slope of $\Delta k$ at the chosen central wavelengths, compared with the slope at shorter wavelengths produced an unusually broad JSA. In this case, the spectral range was capped at $\zeta = 1~\mu$m to avoid undesirable SPDC photon generation far from the central PMF value. Such far-detuned generation is unavoidable in domain-engineered crystals, although it often goes unconsidered as the associated peaks manifest well outside the examined JSA range \cite{graffitti2018design}. In practice these peaks can be ignored due to finite detector bandwidths, motivating the choice to use a capped $\zeta$ value \cite{graffitti2018design}

\renewcommand{\thefigure}{S4}
\begin{figure*}
    \centering
    \includegraphics[width=13cm]{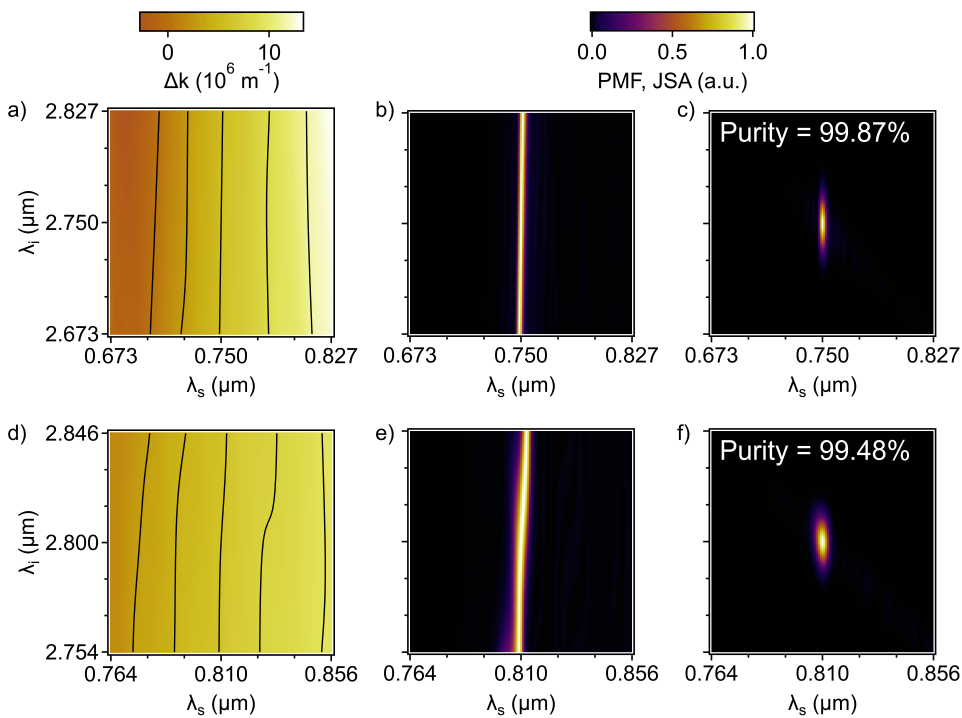}
    \caption{Calculated joint-spectral quantities for both 3R-WS$_2$ and 3R-MoS$_2$ in a non-degenerate wavelength regime. (a) Calculated momentum mismatch $\Delta k(\lambda_s,\lambda_i)$ for 3R-WS$_2$, the angle and linearity of the contours indicates that GVM is achievable. (b) PMF obtained using twist-angle domain engineering. (c) JSA resulting from combining the PMF in (b) with a width optimized PEF, the JSA is near-factorable and yields a spectral purity of $99.87\%$. (d-f) Analogous results as shown in (a-c) using 3R-MoS$_2$, the momentum mismatch in (d) exhibits similar properties to that in (a) leading to the JSA shown in (f) which yields a spectral purity of $99.48\%$.}
    \label{ws2_mos2_results}
\end{figure*}

\section*{Supplementary Material 5: Twist-engineered spdc in realistic, non-centrosymmetric \lowercase{vd}W crystals}

Here, the twist-angle algorithm is applied to two non-centrosymmetric vdW materials with $C_3$ rotation symmetry, 3R-WS$_2$ and 3R-MoS$_2$, using reported material parameters \cite{haastrup2018the, gjerding2021recent}. As in the rBN case discussed in the main text, highly non-degenerate wavelength configurations are considered. Figure \ref{ws2_mos2_results} summarizes the calculated phase-matching and spectral properties of the twist-engineered crystals. Figure \ref{ws2_mos2_results}(a) shows the calculate momentum mismatch $\Delta k(\lambda_s,\lambda_i)$ for 3R-WS$_2$, where the direction and slope of the contours reveal that GVM is achievable. Combining this favorable $\Delta k(\lambda_s,\lambda_i)$ landscape with twist-angle domain engineering using parameters $L = 50\ell_c$, $w = \ell_c = 0.56~\mu$m, and a twist-angle step size of $1^{\circ}$ yields the PMF shown in Fig.~\ref{ws2_mos2_results}(b). The product of this PMF with the corresponding optimized PEF produces the JSA shown in Fig.~\ref{ws2_mos2_results}(c), which yields a high spectral purity of $99.87\%$. 

Analogous calculations were also performed for 3R-MoS$_2$, with the results shown in Fig. \ref{ws2_mos2_results}(d-f). GVM was also found to be achievable in 3R-MOS$_2$, yielding a calculated spectral purity of $99.48\%$. The same twist-engineering parameters as those used for 3R-WS$_2$ were employed, with the exception of the coherence length, which changes to $\ell_c = 0.51~\mu$m.

Together with the rBN results presented in the main text, these findings demonstrate that twist-angle domain engineering is both applicable and effective in real vdW crystals, and that it enables the generation of spectrally pure heralded photons in wavelength regimes characterized by large signal–idler separation.

\section*{Supplementary Material 6: Non-Gaussian nonlinearity profiles}

Although this paper focuses on generating pure photons, the twist-angle algorithm’s precise nonlinearity engineering has broader applications. For generating time-frequency entangled modes, the target PMF can be a Hermite-Gauss (HG) function \cite{graffitti2020direct}, which requires high-precision nonlinearity engineering for accurate reconstruction. Figure \ref{HG_modes} shows the nonlinearity profiles and PMFs for the first- and second-order HG modes, alongside approximations calculated using the sub-$\ell_c$ domain and twist-angle algorithms (with $w=\ell_c$ and a twist-angle step size of $1^\circ$). The average cost function values for the $\phi_{\mathrm{track}}(z)$ approximations are reported in Table \ref{table1} as a basic measure of approximation quality. For brevity, a full analysis is omitted; this section serves to demonstrate the twist-angle algorithm’s ability to precisely reconstruct a wide range of PMFs with superior accuracy compared to current methods.

\begin{table}[h!]
    \centering
    \begin{tabular}{|c|c|c|}
         \hline
         \multicolumn{3}{|c|}{Average cost function values (a.u.)} \\
         \hline
         \hline
         & Sub-$\ell_c$ domains & Twist-angle\\
         \hline
         First-order HG &  0.00993 &  0.00016\\
         \hline
         Second-order HG & 0.01020 & 0.00025\\
         \hline
    \end{tabular}
    \caption{Average cost function values for different target PMFs and domain-engineering algorithms. The twist-angle algorithm consistently yields lower (better) cost function values than the sub-$\ell_c$ domains algorithm.}
    \label{table1}
\end{table}

\renewcommand{\thefigure}{S5}
\begin{figure*}
    \centering
    \includegraphics[width=17cm]{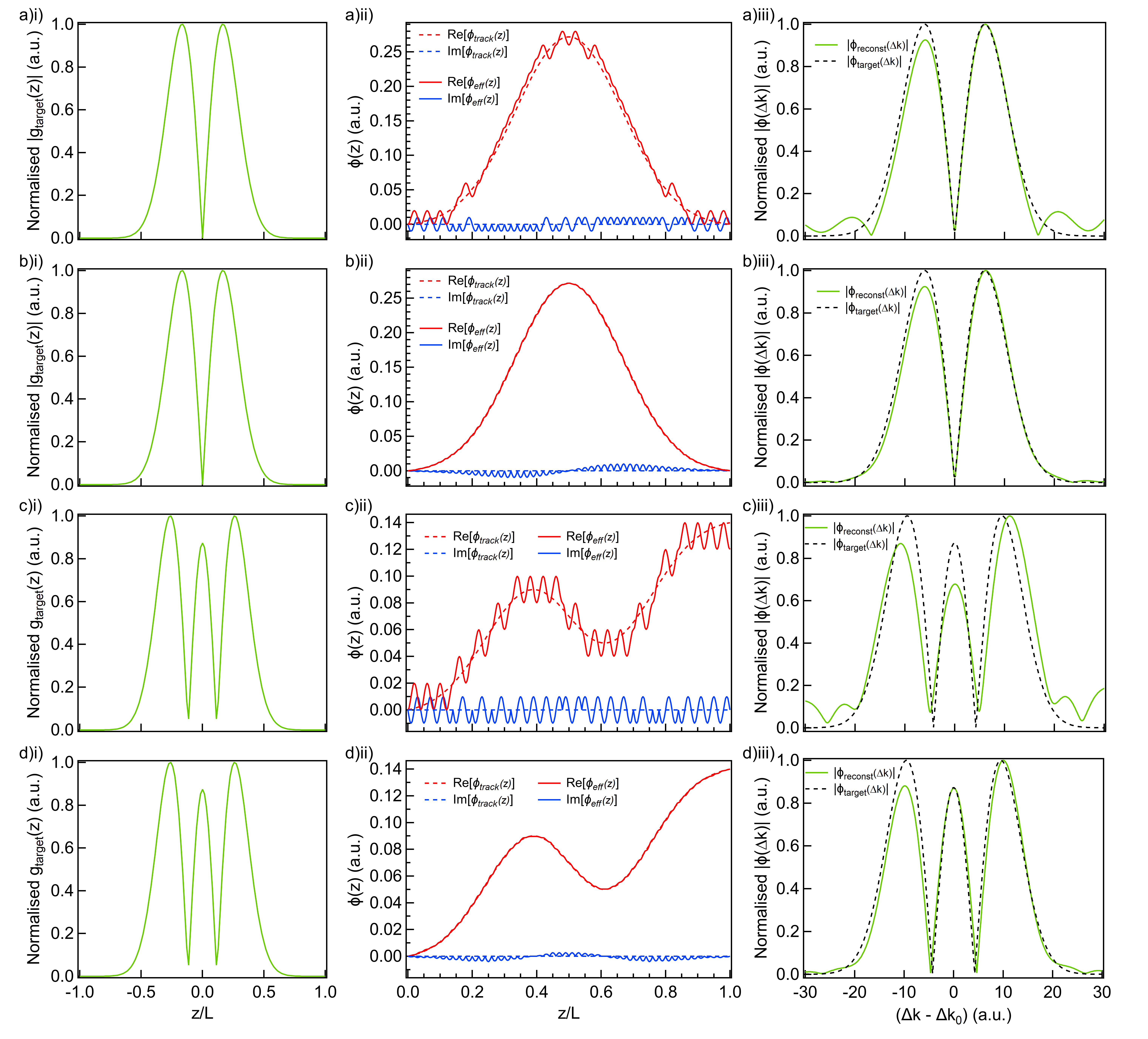}
    \caption{Comparison of the twist-angle and sub-$\ell_c$ domain algorithms in approximating Hermite-Gauss (HG) shaped PMFs. a)i–iii) First-order HG mode: (i) Normalised nonlinearity profile for a crystal centred at $z=0$; (ii) PMF longitudinally through the crystal $\phi_{\mathrm{track}}(z)$ and approximation $\phi_{\mathrm{eff}}(z)$ calculated using the sub-$\ell_c$ domains algorithm; (iii) corresponding normalised PMF at the end of the crystal $\phi_{\mathrm{target}}(\Delta k)$ and reconstructed PMF $\phi_{\mathrm{reconst}}(\Delta k)$. b)i–iii) Same as a)i–iii), but using the twist-angle algorithm. c)i–iii) Second-order HG mode: (i) nonlinearity profile; (ii) tracking and effective PMFs as a function of $z$; (iii) target and reconstructed PMFs using the sub-$\ell_c$ domains algorithm. d)i–iii) Same as c)i–iii), but using the twist-angle algorithm. In all cases, $w = \ell_c$, and real, imaginary, and absolute values are plotted in red, blue, and green, respectively.}
    \label{HG_modes}
\end{figure*}

\end{document}